# A Platform for Teaching Applied Distributed Software Development

## The Ongoing Journey of the Helsinki Software Factory

Fabian Fagerholm*, Nilay Oza†, Jürgen Münch‡
Department of Computer Science, University of Helsinki
P.O. Box 68 (Gustaf Hällströmin katu 2b)
FI-00014, Finland
*fabian.fagerholm@helsinki.fi, †nilay.oza@cs.helsinki.fi, ‡juergen.muench@cs.helsinki.fi

*Abstract*—Teaching distributed software development (DSD) in project courses where student teams are geographically distributed promises several benefits. One main benefit is that in contrast to traditional classroom courses, students can experience the effects of distribution and the mechanisms for coping with distribution by themselves, therefore understanding their relevance for software development. They can thus learn to take more care of distribution challenges and risks when starting to develop software in industry. However, providing a sustainable environment for such project courses is difficult. A development environment is needed that can connect to different distributed teams and an ongoing routine to conduct such courses needs to be established. This article sketches a picture of the Software Factory, a platform that supports teaching distributed student projects and that has now been operational for more than three years. We describe the basic steps of conducting Software Factory projects, and portray experiences from past factory projects. In addition, we provide a short overview of related approaches and future activities.

## I. INTRODUCTION

The Software Factory is an experimental laboratory that provides an environment for research and education in software engineering, and that was established by the Department of Computer Science at the University of Helsinki [1]. Since the first project in 2010, the Software Factory has been used as a platform for teaching software engineering in close collaboration with industry. The goal is to provide students with a realistic environment in which to integrate previous knowledge of computer science and software engineering with experiential insights about conducting real software projects. Close customer involvement, intensive teamwork, and the use of modern software development tools and processes add realism and working life 0,,relevance for the students.

The Software Factory's particular educational focus is teaching global software engineering. Students benefit by learning particular skills that are relevant to globally distributed software development. The Software Factory concept can and has been replicated in other locations, forming a growing network for research and education.

Although the Software Factory concept started initially with collocated projects, the Software Factory aims to conduct mainly distributed projects. Today, several companies and universities have established the Software Factory concept on their premises and several distributed projects have been performed (e.g., a pilot project among Finnish universities, a project with Spain-based Indra Sistemas and Technical University of Madrid on intelligent power grids, and a large-scale open source collaboration project with Stanford University, Facebook, and other academic partners worldwide).

## II. EXPERIENTIAL AND PROJECT-BASED LEARNING IN SOFTWARE ENGINEERING

Teaching DSD can be done in many different ways. Classical classroom teaching is usually limited to transferring knowledge about methods or techniques and conducting exercises. Typically, the exercise examples are unrealistically small, and it is difficult to show the complexity of distributed projects in such settings. In the context of real software development projects, students can experience the effects of such methods and techniques (e.g., the risk of making wrong assumptions without appropriately documented code) and see their practical relevance. Studies have shown that such experiences can also lead to performance improvements on the individual as well as the team level [2]. Involving students as subjects in experiments (e.g., [3]) or using simulators for teaching are other alternatives that allow students to experience or explore the effects of software development techniques.

In the area of DSD, several efforts have been made to provide realistic project environments for distributed student projects:

The Siemens Global Studio Project [4] was one of the first projects that aimed at learning from the collaborative development of student teams in a distributed project environment. In contrast to the Software Factory, the focus of the Global Studio Project was on conducting the project itself, whereas the Software Factory focuses on the development environment and the respective processes needed to operate the environment. The Software Factory has shorter project cycles and faster iterations than the Global Studio Project.

The DOSE course [5] embeds a distributed project in an overall course on teaching distributed development. Compared to the Software Factory, DOSE puts less emphasis on the laboratory setting but has a stronger focus on teaching specific techniques, such as API design.

Several other approaches and frameworks for teaching DSD via student project settings have been reported in the literature. Damian, for instance, describes a framework that uses Scrum practices to teach distributed development [6]. Fortaleza et al. provide a comparison of 19 global software engineering courses [7]. Student and teacher experiences with distributed development courses have been reported in many different ways (see, for instance, [8]). All these reports have influenced the design of the Software Factory approach, as described subsequently.

## III. THE SOFTWARE FACTORY APPROACH

A Software Factory project is an advanced master's-level capstone project course at the Department of Computer Science, University of Helsinki. Student participants work in the Software Factory facility for an average of six hours per working day, and can choose between four or five working days per week. Projects last seven weeks and use agile software development methodology to rapidly produce a functionally complete software prototype in cooperation with an external customer. Software Factory projects are conducted in a manner that simulates, as closely as possible, the reality of software development in new product development organizations. The model is either a small software development company or a division of a large corporation. Some projects, however, may include continued development of existing software, and code reuse, e.g., through open source components, is encouraged where applicable. The projects are conducted in a laboratory setting: a standardized but customizable development environment with a specified physical design (i.e., an interior design pattern with specific furniture and equipment, and different activity-related zones), a defined technical infrastructure, and a comprehensive experimental infrastructure that includes instrumentation for performing empirical studies.

From an educational perspective, the Software Factory provides students with a realistic experience that serves to integrate their previous theoretical and practical knowledge with working life relevance in order to develop higher-order skills. Students are expected to take responsibility for the entire project, including project management, customer communication, iterative requirements solicitation, continuous development process improvement, and, naturally, the software development itself. The approach also allows teachers to supplement the Software Factory projects with other courses on top of the Software Factory activities; examples include courses on software project leadership, project management, group dynamics, software architecture, and software processes, as well as intensive courses on technical topics, such as version control, programming languages, and testing.

While Software Factory projects include much of the uncertainty and open-endedness of real software projects, an established driving process is always in place to provide a frame within which the projects are conducted and learning can occur. The process can be divided into pre-project, per-project, and post-project stages. The different stages of this process are described subsequently.

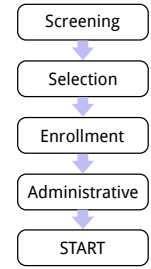

Fig. 1. Process for pre-project activities.

### A. Pre-project activities

The pre-project activities aim to reach a defined state in which the project can be handed over to the implementation team. The project's prerequisites must be fulfilled, but only to the extent necessary for the implementation team to take over. In distributed projects, pre-project activities are especially important. Once a project has started, it is often not feasible to make major changes to schedules and allocation of personnel resources. At the same time, the exact details often depend on experiences gained during the start of the project. An overview of the pre-project process is shown in Figure 1.

*1) Project screening and selection:* From the project perspective, the most important pre-project activity is the selection of a project idea and the initial work that prepares the customer to interact with the implementation team. This activity can be considered as a project portfolio management task. Proposals may arrive through multiple means, including direct contact with the Software Factory staff or through an online project proposal form on the Software Factory web site[1]. The first step is screening, where the minimum prerequisites are evaluated and feedback is given directly to the potential partner. This step is continuously carried out as proposals arrive. Selection proceeds by considering project proposals that have passed the screening stage. Here, the Software Factory works much as a large software development organization: projects are considered in terms of their feasibility, maturity, and contribution to organizational goals, which in this case stem from both educational and research needs. Proposals are either accepted or postponed. In the case of acceptance, the partner is asked to produce more detailed material to be used as the first high-level requirements description. Postponed projects are reconsidered for the following cycle, and the partners may update or withdraw their proposals. A particular challenge in this phase is how to screen project proposals where customers are remotely located. The ability of the customer to communicate in such an environment is of critical importance.

---

[1]http://www.softwarefactory.cc/

*2) Enrollment:* An enrollment stage is used to screen students before admission. During this stage, minimum admission requirements are checked. Students are assessed so that a selection can be made in case the number of eligible applicants exceeds the project's capacity. Another objective of this assessment is to match the students' skills to the project's known needs. Consequently, a skilled, motivated, and competitive team of 5-15 students is formed. In some cases, where multiple projects has been selected for simultaneous implementation, students may be divided into multiple teams.

*3) Administrative issues:* Once the team composition is known, several administrative issues must be handled. These vary for different universities, but may include things like ordering keys for the Software Factory facility, setting up user accounts for the technical infrastructure. In a global setting, this phase is particularly challenging, with varying processes among university IT departments. These departments are often not prepared to provide services to parties outside their university. Also, local policies may interfere, and student teams must spend time on working around technical and policy issues. A standardized lab environment overcomes many of these problems as remote students and customers can be granted access rights to the systems in a uniform way.

*4) Start: Project kickoff:* We have developed specific project kickoff activities for getting the project team up to the speed and style of real-life software development. We emphasize self-directed learning practices, which help students to realize that they are expected to take initiative and engage in the project. This requires changing the students' mind-set away from the familiar lecturing style, where the initiative is teacher-driven and based on presentations and instructions. Rather, we present the project as an open-ended learning problem where the students must seek the information they need to solve the problem and its parts. Understanding the problem itself and evaluating the solution are parts of the goal. Another aim of the kickoff is to direct the students' attention to the needs of the customer, and the customer's attention to the students as the primary point of contact for getting things done in the project. This ensures that communication between the team and the customer is direct, and that it is initiated spontaneously when either party observes a need.

With multiple teams in different locations, there are a number of options available for conducting the kickoff. In practice, we have had the most positive experiences by arranging a co-located event at the start of the project. This is also an important lesson to learn: meeting in person can reduce the barriers for continued communication online. When this is not possible, teachers may want to carry out the activities using online tools. In this case, it is important that each site has a local instructor who facilitates communication and encourages students to engage with their remote team mates and not only with the co-located ones.

In practice, the exact implementation of the kickoff activity can vary, but it always has the following three elements, which are based on the Extreme Apprenticeship (XA) method and its three stages of modeling, scaffolding, and fading [9].

*Modeling* is grounded in material that the customer brings to the project. This includes verbal descriptions, diagrams, written documents, and any other material that the customer chooses. Administrative material provides the organizational constraints for the project. The teacher provides the necessary *scaffolding* by directing how the material is processed. The activity proceeds from individuals to the whole group. Tasks are given first to individuals or pairs, and then gradually, larger subgroups work on larger parts of the problem space. Finally, the team and customer representatives work as a single group to define a first sketch of the whole project. The teacher's involvement decreases during the process. In the beginning, explicit instructions are given. Gradually, the teacher *fades* into the background until he or she only provides support and instruction when asked. At the very end of the activity, the teacher encourages the students to work in a similar way throughout the project. The teacher also solidifies all of the participants' beliefs that they can reach a meaningful outcome for the project. Perhaps most importantly, the teacher explicitly transfers responsibility of the project to students and customer representatives. The teacher then assumes a supporting role and is available on demand, but can intervene if necessary.

Another common element is the use of agile software development methodology, specifically, the so-called Scrumban process [10], [11]. This process combines many of the Scrum practices with the visual Kanban planning board. The use of this method is gradually introduced, first through an example. However, the value of this method for a project is only visible once there are actual tasks to perform. The process is linked to the previous exercise in order to introduce a systematic element to the cycle of discovery, requirements specification, implementation, and evaluation. In global projects, the physical Kanban board can be replaced by online variants or omitted. A local board may be used to facilitate local work, but some additional effort is then required to synchronize information to remote teams.

*B. Per-project activities*

Since each project varies considerably in the project team, customer, topic, and technology choices, the common per-project activities are fairly general. We have found it beneficial to maintain a regular cycle with weekly customer meetings where the team demonstrates the current state of the software and the customer gives direct feedback to steer the next cycle. Online demonstrations should be well prepared. Screen sharing or other technical means to give the customer an opportunity to conduct interactive demonstrations.

In addition to weekly meetings, we follow the Scrum practice of daily meetings. Team members shortly answer three questions: 1) "What have you done since the last meeting?" 2) "What are you planning to do until the next meeting?" and 3) "Are there any obstacles preventing you from carrying out your work?" However, we also acknowledge that for some project stages and some kinds of tasks, daily reporting may be too frequent, and in these cases, the meetings do not have to follow this exact format as long as it fulfills the spirit of efficient information-sharing. Special care should be taken when holding weekly meetings online. Varying image and audio quality may introduce communication overhead. In our experience, successful online meetings require both a meeting moderator who keeps the pace and structure of the meeting, and on-site technical support to ensure that each participant can hear and see each other. In many cases, it may prove more effective to conduct such meetings over text chat. In any case, our experience shows that a separate local meeting is often needed to discuss more intricate details.

Finally, in order to provide the team with access to relevant information, we invite the customer to be available frequently for free-form discussions with the team. This must be balanced with enough time for the team to focus on actual implementation. Through these interactions, the team can access the customer directly for key decisions, and can learn the skill of iteratively soliciting requirements. Encouraging the customer to be available online regularly is a good way of enabling this free-form communication.

*C. Post-project activities*

Apart from administrative tasks, such as closing accounts, returning room keys, and other such matters, what remains from the educational perspective is to properly debrief project participants in order to engage the whole group in reflection. A summative assessment of the students is also performed.

The debriefing session is conducted differently depending on the events during the project and its outcome. Generally, a good approach is to analyze the project through a time-line, where the students and customer representatives recall the phases of the project chronologically. The teacher asks open-ended questions to encourage the participants to reflect deeply on causes and effects and different interpretations. As with the daily meetings, we find that subtle, but important details may be lost in online communication. Therefore, we always arrange a local debriefing session for our students. Ideally, this event would be co-located, but we have so far not explored this in practice. Finally, summative assessment of experiential learning is a challenge in itself, and is outside the scope of this paper. We note that peer assessment can provide students with an opportunity to reflect on their role in and performance on the project.

## IV. Experience and Results

While there are several challenges involved in conducting Software Factory projects, we find the overall results to be encouraging. By employing a systematic driving process, we have been able to reduce the administrative burden, and have allowed the teachers to focus on the educational aspects of the projects. In this section, we report on particular experiences with specific projects.

*A. Sustainability*

One of the important aspects of an endeavor such as Software Factory is sustainability. Our initial investment in properly planning the overall setup and consulting with all relevant stakeholders, including companies, students, and researchers, helped us to develop a course that has minimal overhead and maximum support from all stakeholders. A particular challenge is how to sustain continued development. This requires strong support from the department as well as funding for personnel and equipment maintenance. We believe that our adaptive approach has been a key factor in both gaining support and utilizing existing funding effectively. Our results with making the environment systematic without making it static show that the financial requirements can be scaled up and down while still keeping the educational value intact.

*B. Globally distributed projects with remote teams, customers and technology*

We have worked with distributed partners including off-site customers, development teams from other factory nodes, and also distributed and remotely located technology infrastructure. Software Factory has helped students and researchers understand new levels of complexity in distribution – in relation to people, technology and processes.

As an example we recently conducted a joint project between Helsinki and Madrid teams (from Technical University of Madrid and Indra Software labs) where the Helsinki team joined an ongoing software product development project. Students gained a unique experience of working with a completely unknown team. They also developed hands-on experience on how to deal with cloud infrastructure, both from a technical perspective and an operational perspective: deciding on access controls, and using a shared code base. Students also gained experience with keeping to a development process. Just deciding to use Kanban was not enough; students had to work quite a lot to better understand, negotiate and fine-tune their approach to task assignment, allocation and commitment.

From our experiences, we can identify a number of challenges with conducting distributed educational projects with other universities. These often stem from the same underlying reasons that make professional distributed projects difficult: distances in time, location, and culture. Complete synchronization of teaching schedules is often impossible. We have attempted to turn this into a learning experience: it is common for distributed projects to have staged starts, with different locations starting at different times. Another challenge is in student selection: each university applies their own prerequisites and standards in student selection, and therefore, there may be differing levels of skill in the different locations. We have chosen to accept this risk and attempt to mitigate it for our students by keeping our own selection baseline high and including handling the overhead of differing skills levels in the project scope. Grading poses a final challenge to overcome, again with different standards at different universities. We have chosen to be inclusive in the grading, utilizing the perspective of several project participants as material for grading.

*C. Team and Student Considerations*

Our projects have relied heavily on our approach to building self-organized teams [12]. Being able to operate in such a team is a learning goal in itself. Relying on self-directed students has also helped us a great deal in coordinating the whole course and keeping stakeholder communication efficient.

A large number of our master's-level students have past industry work experience, which helps in conducting professional software projects. We carefully match the students' technical skills and experience with the needs of the project.

*D. Project Considerations*

All projects in the Software Factory undergo the highly iterative, Scrumban development process in a cross-functional, self-organized team environment. We do not provide a project manager for the students. Instead, as previously noted, the teacher supervises the project, is available on demand, but can intervene if needed. In addition, a resident coach actively mentors the team and makes sure that the project lives up to its expected outcomes for the customer. The coach also frequently engages with the customers to help them interact with the team and focus on the underlying reasons for their wishes and on their choices regarding the next step towards the project goals.

We have to be quite selective regarding which ideas we work on. We tend to select projects that add concrete value to the customer's offerings and try to avoid developing "unusable" prototypes with dubious value. This also encourages our customers to be quite active in their involvement during the project. In our experience, close customer participation is a critical success factor for producing a software product in seven weeks with a newly composed software team. This is a particular challenge when the customer is not collocated with the team. Extra effort is needed to ensure that the customer actively participates in the team's work and provides the feedback necessary for rapid prototyping.

*E. Involving students in research*

Several researchers have utilized the Software Factory platform. Specific lines of inquiry have ranged from examining sources of waste in Scrumban to psychometric analysis of team behavior. We have found that these ongoing research efforts are also interesting for student participants, as they are keen to see results concerning their own projects. In some cases, researchers have been able to provide the project with empirically based real-time feedback on different aspects of the project performance. Also, some students have utilized the platform for empirical studies in their Master's theses. We believe there are many opportunities for teaching empirical research skills in the context of the Software Factory.

## V. NEXT STEPS

There are several ongoing developments to evolve the Software Factory concept. One is to use cloud services for development, management, and coordination of the projects. This concept is referred to as the "Cloud Software Factory" and has been partly implemented in Helsinki.

Several organizations have networked and already established or are establishing Software Factories. Current sites are in Helsinki, Oulu, and Joensuu (Finland), Bolzano (Italy), and two in Madrid (Spain). Sites are planned in Ostrava (Czech Republic) and Novi Sad (Serbia). There are also prospects for sites in China, and collaboration with other European and North American universities.

A future direction is to integrate empirical studies more systematically into Software Factory projects. The relatively short setup times for projects in the Software Factory currently make the planning of accompanying empirical studies difficult. However, several longitudinal studies are currently being conducted that are less sensitive to individual project schedules.

Finally, customers could be involved in a wider sense. Software Factory can serve to support experimentation with customer value. During the course of a project, prototypes (minimally viable products) would be developed and used by the customer to directly test value with end users. Based on experimental results, development goals might be adjusted.